\documentclass[aps,pre,showpacs,showkeys,twocolumn,groupedaddress]{revtex4}

\usepackage{amsfonts,amssymb}
\usepackage{graphicx}
\usepackage{mathbbol}

\newcommand {\tr} {{\rm tr}}

\begin{document}

\title{Parametric correlations versus fidelity decay: the symmetry breaking case}



\author{H. Kohler$^{1}$\ and T. Nagao$^2$ \ and H.-J. St\"ockmann$^3$}
\email{hkohler@icmm.csic.es}
\email{nagao@math.nagoya-u.ac.jp}
\email{stoeckmann@physik.uni-marburg.de}
\affiliation{ $^1$ Instituto de Ciencia de Materiales de Madrid, CSIC,
Sor Juana de la Cruz 3, Cantoblanco, 28049 Madrid, Spain\\
$^2$Graduate School of Mathematics, Nagoya University,  
Chikusa-ku, Nagoya 464-8602, Japan \\
         $^3$ Fachbereich Physik der Philipps-Universit\"at
Marburg, Renthof 5, D-35032 Marburg, Germany\\
}

\date{\today}

\begin{abstract}
We provide formulae for fidelity decay and parametric energy correlations for random matrix ensembles where 
time--reversal invariance of the original Hamiltonian is broken by the perturbation. Like in the case of a symmetry conserving 
perturbation a simple relation between both quantities exists. Fidelity freeze is observed for systems with even and odd spin.
\end{abstract}

\pacs{05.45.Mt,03.65.Yz,03.67.Lx}
\keywords{Fidelity, Parametric energy correlations, Random matrix theory}

\maketitle

\section{Introduction}

Fidelity  presently attracts considerable attention in diverse fields like quantum information, 
quantum chaotic systems and others
\cite{gor06,jac09}. It measures the change of quantum dynamics of
a state under a modification of the Hamiltonian. In quantum
information, fidelity measures the deviation between a
mathematical algorithm  and its physical implementation.

Since fidelity requires knowledge of the entire wave function for the original and for the modified 
system a measurement of fidelity is a notoriously difficult task. 
However a number of experimental results have been obtained in microwave 
billiards, where the perturbation was achieved by varying some geometric 
parameter. There are two qualitatively different ways to do this, either 
by a global perturbation, e.g. by moving one wall \cite{schae05b}, or a local 
perturbation, e.g. by varying the position of an impurity \cite{hoe08}. 
For the first  case random matrix theory is applicable, and indeed a 
perfect agreement between  experiment and  theory was found \cite{schae05b}. 

On the other hand statistical properties of energy correlations between spectra of complex quantum 
systems which differ by a parameter-dependent variation have been studied experimentally and 
theoretically~\cite{genpar}. This quantity can be obtained with great accuracy from scattering 
experiments by analyzing the fluctuations of the resonances in the scattering cross--section \cite
{ber99,hul09}.  

From an experimental point of view it is interesting to relate fidelity with spectral quantities. This 
allows an indirect measurement of fidelity via  an analysis of the (parametric) scattering data and 
the problem of measuring the entire wave function is circumvented.

A simple differential relation between fidelity decay and parametric energy correlations was 
established in the case that the parameter dependent perturbation falls into the same symmetry 
class as the unperturbed system \cite{koh08,smo08}.  This differential relation was derived earlier 
in energy space by Taniguchi, Simons and coworkers \cite{sim95,tan96} and it was identified with a 
continuity equation of the Calogero--Moser--Sutherland model \cite{tan95}.  In Ref.~\cite{tan94}
similar expression were derived for parametric energy correlations in the case where the perturbation breaks 
the global symmetry of the orgininal unperturbed system.

Recently billiard experiments could be performed in microwave resonators, where time reversal 
symmetry (TRS) was broken by a piece of ferrite \cite{die09} which plays the role of the 
perturbation. From the experimental results $S$--matrix elements could be determined and an 
estimate of the strength of the TRS--breaking was made.  The experimental setup seems adequate 
 for a measurement of parametric energy correlations and of fidelity decay by a TRS 
breaking perturbation.  

In this paper we therefore analyze the expressions found in Refs~\cite{tan94,sim95} for TRS--breaking perturbations 
under the aspect of fidelity and provide formulae for fidelity and parametric form factor as well as differential relations 
between them and discuss their consequences.

\section{Definitions and Results}
\label{mainresults}
Fidelity amplitude is defined as a functional of the initial wave function. In an ergodic 
situation it seems reasonable to replace a specific initial state by a random one. In Ref.\cite{gor04} the 
corresponding random matrix model for the fidelity amplitude 
was defined by ($\hbar = 1$)
\begin{equation}\label{ss01c}
  f(\lambda_\parallel,\lambda_\perp,t)\ =\ \frac{1}{N}\left<\tr \exp(i tH )\exp(-i tH_0)\right>\, .
\end{equation}
The Fourier transform of parametric energy correlations is defined by 
\begin{equation}\label{ss011a}
  \widetilde{K}(\lambda_\parallel,\lambda_\perp,t)\ =\ 
           \frac{1}{N}\left<\tr \exp(i tH )\tr \exp(-i tH_0 )\right>\, .
\end{equation}
It was named cross form--factor in \cite{koh08}.  The brackets denote an ensemble average.  The 
perturbed Hamiltonian $H$  is given as
\begin{equation}\label{ss01a}
 H \ =\ H_0 + \lambda_\parallel V_{\parallel} + \lambda_\perp V_\perp  .
\end{equation}
Let us first discuss the unperturbed Hamiltonian. We assume that for the unperturbed system $H_0$ TRS is conserved. 
The time reversal operator ${\cal T}$ acts differently on systems with integer spin and on systems with half--integer 
spin \cite{mes99}.  For even spin ${\cal T}_1 = \hat{C}$, where $\hat{C}$ is the complex 
conjugation operator. In this case (called case I in the following) $H_0$ is chosen from the 
ensemble of real symmetric matrices, called the Gaussian orthogonal ensemble (GOE, $\beta=1$).  
For odd spin systems  ${\cal T}_2 = \hat{C}\hat{J}$, where $\hat{J}$ acts via conjugation with the 
symplectic metric. In this case (case II) $H_0$ is chosen from the Gaussian symplectic ensemble 
(GSE, $\beta=4$), consisting of all Hermitean matrices which are invariant under ${\cal T}_2$. The 
ensembles are defined by the averages 
\begin{equation}\label{ss03}
    \left<(H_0)_{ij}(H_0)_{kl}\right> \ =\ \left\{\begin{array}{cc}
     {\displaystyle\frac{N}{\pi^2}}\left(\delta_{il}\delta_{jk}+\delta_{ik}\delta_{jl}\right)\, &
      {\rm GOE}\\[0.5em]
     {\displaystyle\frac{N}{\pi^2}} \left(\delta_{il}\delta_{jk}-\frac{1}{2}\delta_{ik}\delta_{jl}\right)\, & {\rm 
GSE\ .}\end{array} \right.
\end{equation}
For the GSE the matrix entries are quaternions. Zirnbauer and Altland classified random matrix 
ensembles along Cartan's classification of symmetric spaces with curvature zero \cite{alt97,zir10}. 
They called the GOE of AI--type and the GSE of AII--type. 

In contrast to Ref.~\cite{sto06} we now
assume that the perturbation contains two parts. One part, named $V_\parallel$, shares the 
symmetry of $H_0$ and is taken either from a GOE or a GSE  
\begin{eqnarray}\label{ss06}
    \left<(V_\parallel)_{ij}(V_\parallel)_{kl}\right> \ =\ \left\{\begin{array}{cc} \delta_{il}\delta_{jk}+
\delta_{ik}\delta_{jl}\, & {\rm GOE}\\[0.2em]
        \delta_{il}\delta_{jk}-\frac{1}{2}\delta_{ik}\delta_{jl}\, & {\rm GSE\ .}\end{array} \right. \end
{eqnarray}
 The second part, $V_\perp$, breaks the symmetry of $H_0$ and is taken for case I from the 
Gaussian ensemble of matrices which change sign under time reversal ${\cal T}_1$, being 
antisymmetric matrices with purely imaginary entries. The ensemble is of type B in the classification 
of \cite{zir10}. For case II it is taken from the ensemble of matrices which change sign under ${\cal 
T}_2$. They are of the block form
\begin{equation}
\label{Vpargse}
V_\perp = \left[\begin{array}{c|c} A&B\cr\hline B^\dagger& -A^*\end{array}\right] \ , \ A=A^\dagger\ , \ B= B^T \ ,
\end{equation}
where $A$ and $B$ are $N/2 \times N/2$ matrices ($N$ even). 
The corresponding ensemble is termed C--type in  \cite{zir10}. The B--type and C--type 
ensembles are defined by the averages
 \begin{eqnarray}
   \left<(V_\perp)_{ij}(V_\perp)_{kl}\right> \ =\ \left\{\begin{array}{cc} \delta_{il}\delta_{jk}-\delta_{ik}
\delta_{jl}\, & \mbox{\rm B--type}\\[0.2em]
        \delta_{il}\delta_{jk}+\frac{1}{2}\delta_{ik}\delta_{jl}\, & \mbox{\rm C--type\ .}\end{array} \right. 
\end{eqnarray}
The variance of the matrix elements has been
chosen to have a mean level spacing one,  $D=1$, for $H_0$ and to be of order
$1/\sqrt{N}$ for $V_\parallel$ and $V_\perp$. 

One parameter $\lambda_\parallel$ characterizes the strength of the perturbation, which conserves 
TRS. The second one, $\lambda_\perp$, is the strength of the TRS breaking perturbation.  Thereby 
we consider a much wider class of TRS breaking Hamiltonians as before. 
Observe that for $\lambda_\parallel=\lambda_\perp$ this corresponds to a perturbation by a 
Hermitean matrix, i.~e.~to a perturbation which is taken from the Gaussian unitary ensemble 
(GUE). This ensemble is called type A in \cite{zir10}. Thus time reversal symmetry breaking can 
occur in different ways. Symbolically we may write the l.~h.~s. of equation (\ref{ss01a}) as ${\rm AI}
+ \lambda_\parallel {\rm AI} + \lambda_\perp {\rm B}$ (case I) or as ${\rm AII}
+ \lambda_\parallel {\rm AI} + \lambda_\perp {\rm C}$ (case II). Usually only the transition ${\rm AI} + 
\lambda {\rm A}$ is considered, when time--reversal invariance is discussed \cite{guh98}. 

Analysing equation (4) and the following ones of Ref.~\cite{tan94}, we find expressions for 
fidelity amplitude and for cross--form factor. To  present them concisely we define for case I the 
function
\begin{eqnarray}\label{mres1}
 && {\cal Z}^{\rm (I)}(\lambda_\parallel,\lambda_\perp,\tau)\ = \ \int\limits_{{\rm Max}(0,\tau-1)}^\tau 
du\int\limits_0^u v\,dv\nonumber\\	
 &&\qquad \frac{1+4\pi^2\lambda^2_\perp(\tau^2-v^2)}{\sqrt{[u^2-v^2][(u+1)^2-v^2]}}\frac{(\tau-u)(1-
\tau+u)}{(v^2-\tau^2)^2}\nonumber\\
  &&\qquad \qquad e^{-2\pi^2(\lambda^2_\parallel+\lambda^2_\perp)\tau(2u+1-\tau)
          -2\pi^2(\lambda^2_\parallel-\lambda^2_\perp)v^2}\,,
\end{eqnarray}
and for the case II the function
\begin{eqnarray}\label{mres1a}
&&{\cal Z}^{\rm (II)}(\lambda_\parallel,\lambda_\perp,\tau)\  = \ \int_{-1}^{+1}du\int_0^{1-|u|}\frac{(u
+t)^2-1}{(t^2-v^2)^2} 
\nonumber\\
&&\qquad\theta(u-1+t)\frac{v dv\left(1+\pi^2\lambda^2_\perp(\tau^2-v^2)\right)}{\sqrt{\left[(u-1)^2-
v^2\right]\left[(u+1)^2-v^2\right]}}\nonumber\\
& &\qquad e^{-\pi^2(\lambda^2_\parallel+\lambda^2_\perp)\tau(2u+\tau)
          -\pi^2(\lambda^2_\parallel-\lambda^2_\perp)v^2} \ ,
\end{eqnarray}
where $\tau$ is time measured in units of Heisenberg time $t_H=2\pi/D$. Then in the large $N
$--limit the fidelity as defined in Eq.~(\ref{ss01c}) is given in both cases by
\begin{eqnarray}\label{mres2}
  f(\lambda_\parallel,\lambda_\perp,\tau)&=&-\frac{1}{\pi^2}\frac{\partial}{\partial (\lambda_
\parallel^2)}{\cal Z}(\lambda_\parallel,\lambda_\perp,\tau)\ .
\end{eqnarray}
The cross form--factor is given by
\begin{eqnarray}\label{mres3}
  \widetilde{K}(\lambda_\parallel,\lambda_\perp,\tau)&=&\frac{4}{\beta}\tau^2 {\cal Z}(\lambda_
\parallel,\lambda_\perp,\tau)\ .
\end{eqnarray}
From this follows the relation between fidelity and cross form--factor \cite{sim95}
\begin{eqnarray}\label{mres4}
   f(\lambda_\parallel,\lambda_\perp,\tau)&=&-\frac{\beta}{4\pi^2\tau^2}
                \frac{\partial}{\partial (\lambda_\parallel^2)} \widetilde{K}(\lambda_\parallel,\lambda_\perp,
\tau)\ .
\end{eqnarray}
This relation can be derived through  an universality 
argument without going through a lengthy supersymmetric calculation and comparing results.
In appendix \ref{appB} we present this derivation extending the method of Refs.~\cite{koh08,smo08} to the
case of TRS breaking.


Some details on the the derivation of equations (\ref{mres1}) to (\ref{mres4}) from the pertinent formulae of 
Ref.~\cite{tan94} are given in appendix \ref{appA}.

\section{Discussion}

The double integrals (\ref{mres1}) and (\ref{mres1a}) can be evaluated numerically (see appendix B of Ref.~\cite{sto04} for a convenient parametrization).
\begin{figure}
\includegraphics[width=1\linewidth,height=5cm]{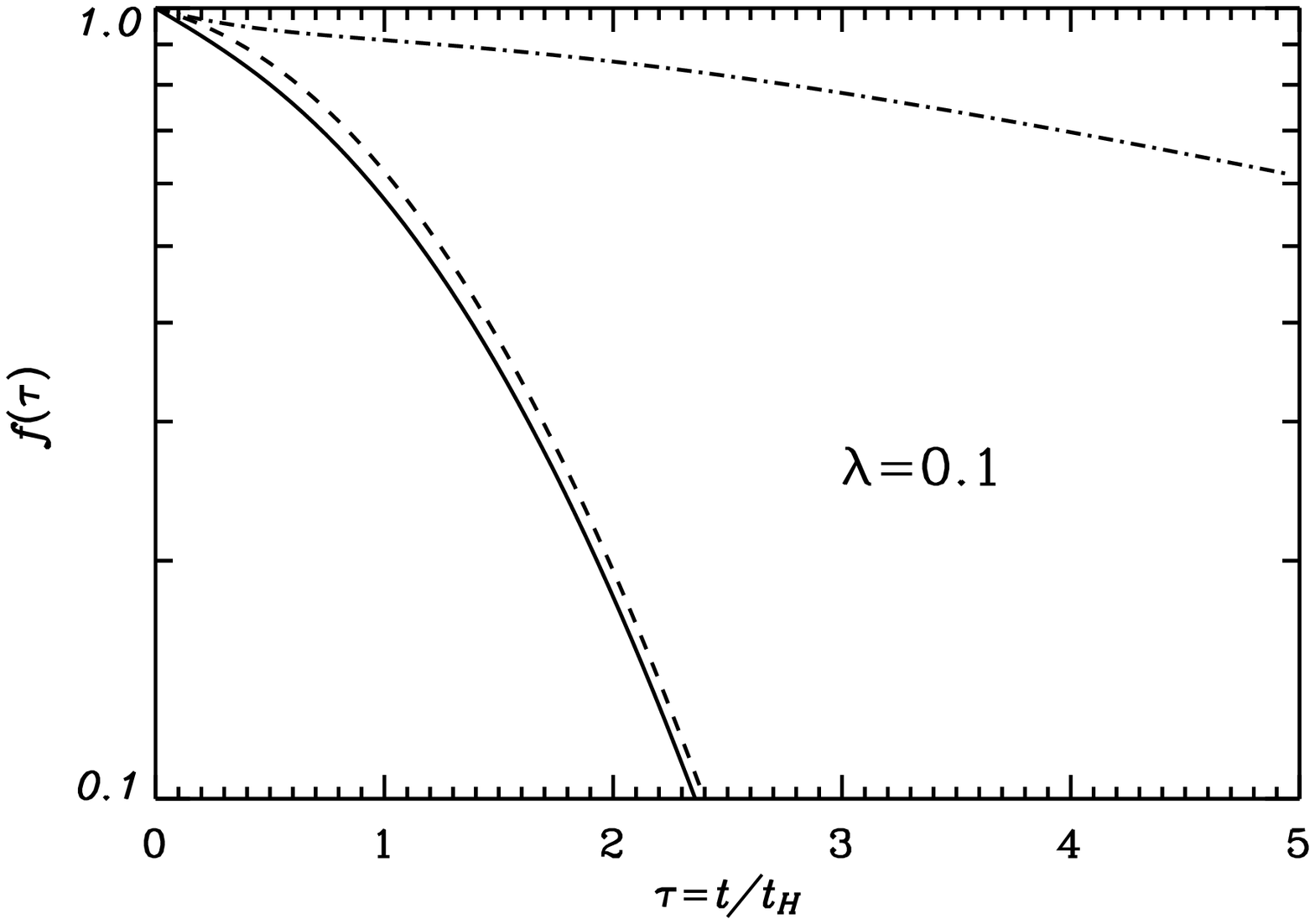}\\
\includegraphics[width=1\linewidth,height=5cm]{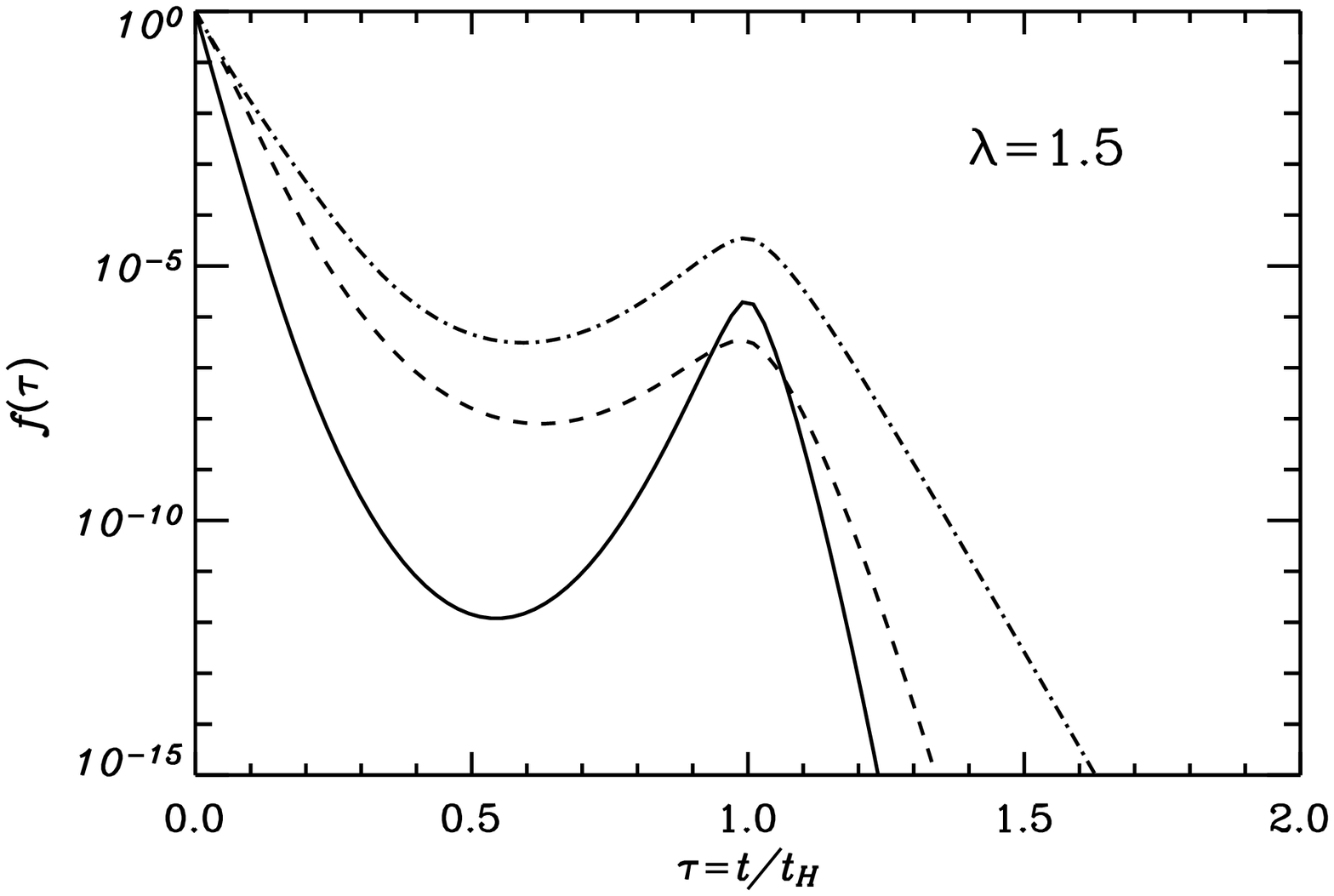}\\
  \caption{Ensemble average of the fidelity amplitude $f(\lambda,\lambda,\tau)$ (solid line) with 
$H_0$ taken from the GOE (case I) and the
  perturbation taken from the GUE for different perturbation strengths $\lambda$.
  The results $f(\lambda,0,\tau)$ (a pure GOE perturbation, dashed lines) and $f(0,\lambda,\tau)$ (a 
purely perpendicular perturbation, dashed-dotted lines) are shown as well for the same parameter.}
  \label{fig:fid1}
\end{figure}
Figure \ref{fig:fid1} shows the fidelity decay in case I for different
perturbation strengths and for a perturbation taken from a GUE 
(${\rm AI}+\lambda{\rm A}$), from a GOE (${\rm AI}+\lambda_\parallel{\rm AI}$) and from the B--type ensemble of purely antisymmetric matrices (${\rm AI}+\lambda_\perp {\rm B}$). For weak
perturbations there is nearly no difference between the fidelity
decay with a GOE and a GUE perturbation, respectively. This feature is closely related with the
extremely weak fidelity decay for an imaginary antisymmetric perturbation. The latter was called 
fidelity freeze \cite{pro05} and was discussed in the context with random matrix theory in 
Ref.~\cite{sto06}. The diagonal elements of the perturbation in the eigenbasis of the original Hamiltonian cause a Gaussian decay, which dominates for times larger than Heisenberg time. It was therefore predicted \cite{pro05} 
that fidelity decay is much slower for perturbations which are purely off--diagonal in the eigenbasis of the original Hamiltonian.

With increasing perturbation the decays for the GOE and the GUE
perturbation separate, and the freeze behavior get lost. For strong perturbations a recovery of 
fidelity at Heisenberg time is seen. This is already known from  \cite{sto05} where the cases 
${\rm A}+\lambda{\rm A}$ and ${\rm AI}+\lambda{\rm AI}$ were discussed. 

For small perturbations and for times much smaller than Heisenberg time fidelity decay is governed by Fermi's golden rule. In this regime the crucial parameter is $\lambda^2=\lambda_\parallel^2+\lambda_\perp^2$ which is related to the spreading width $\Gamma=2\pi \lambda^2 D$ of an unperturbed state.  This result holds independently of the universality class of the background. It is therefore interesting to look on the fidelity amplitude for fixed $\lambda$ but different ratios between orthogonal and parallel perturbation.

\begin{figure}
\includegraphics[width=1\linewidth,height=5cm]{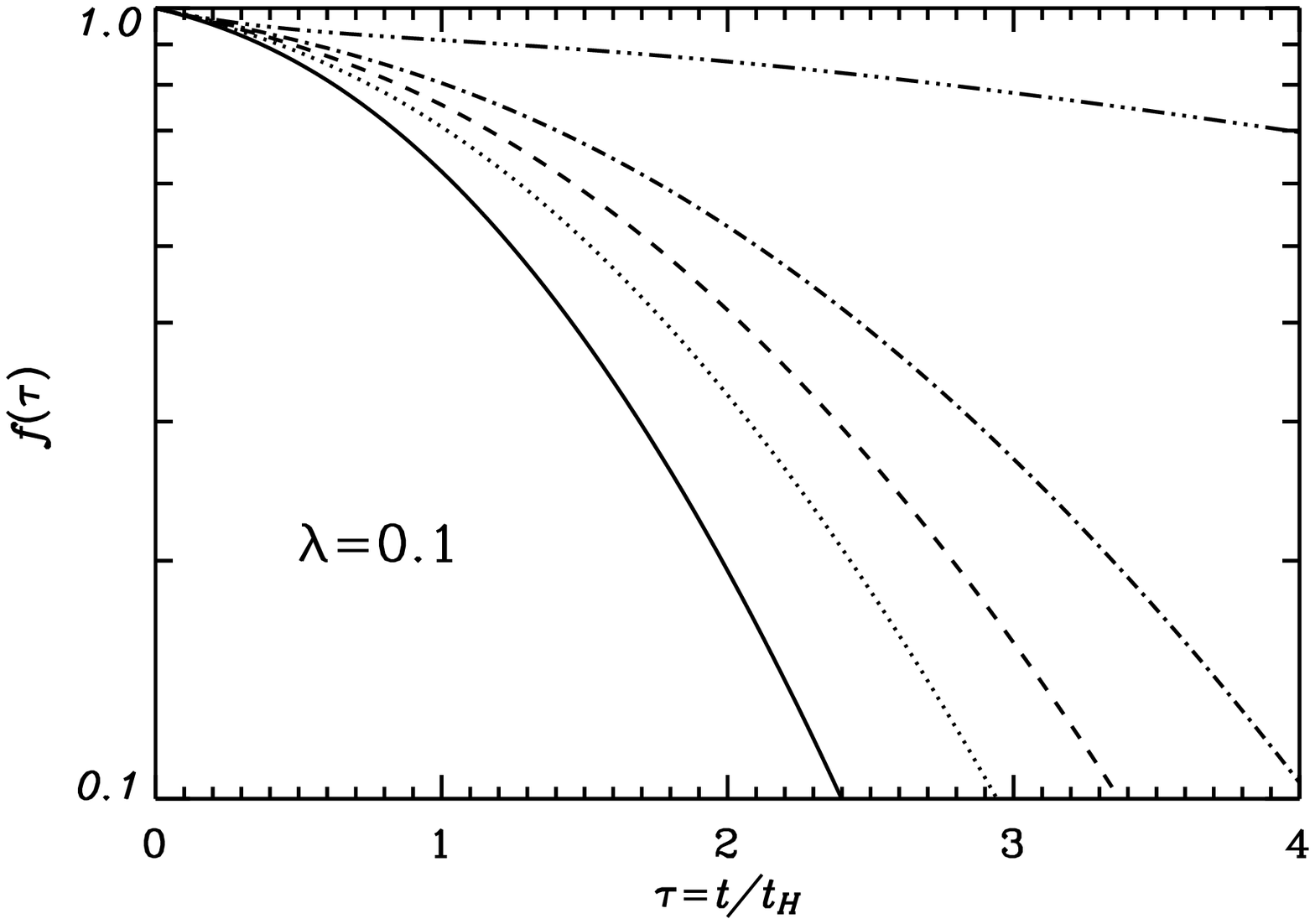}\hfill
\includegraphics[width=1\linewidth,height=5cm]{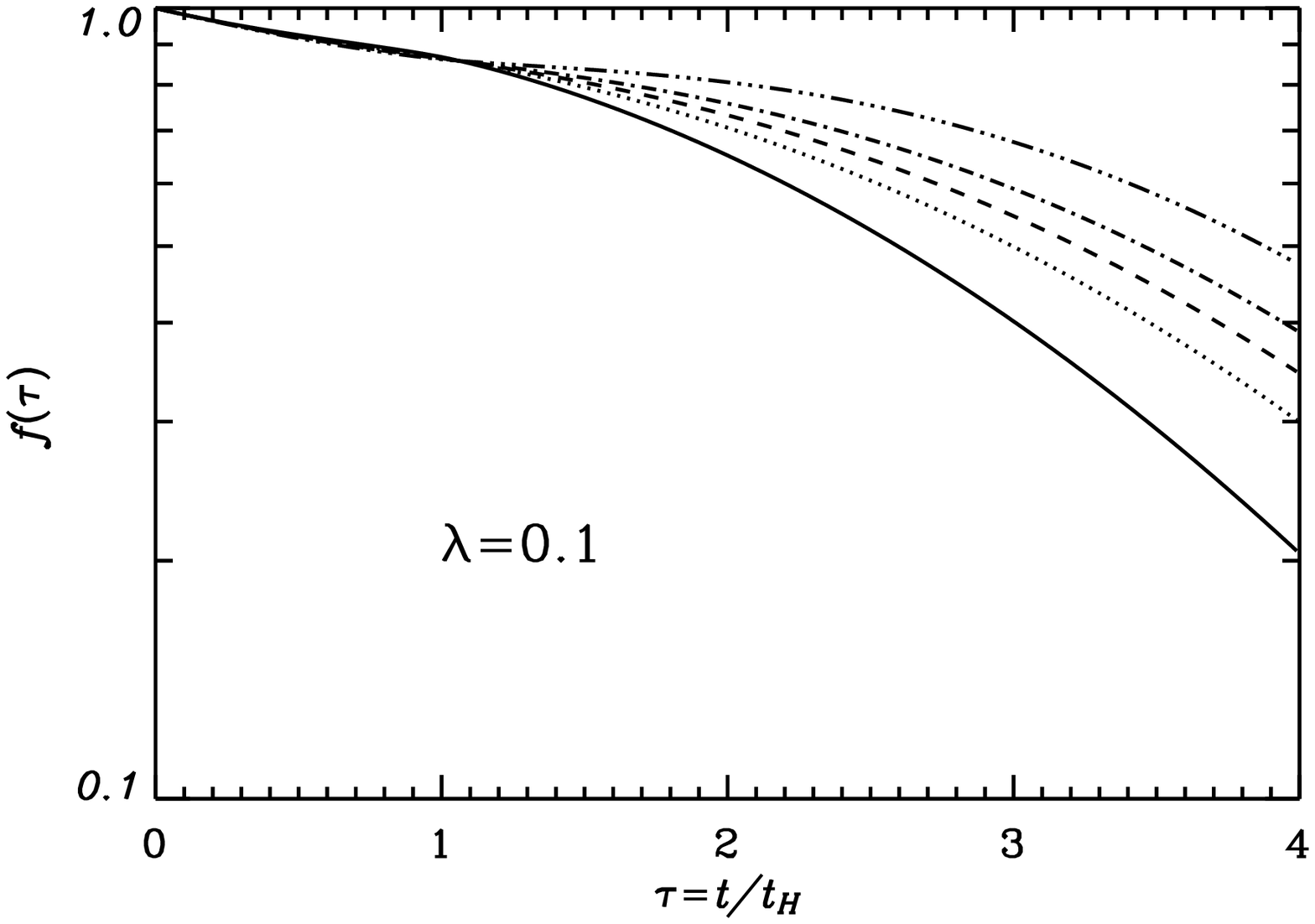}\\
  \caption{ Fidelity amplitude $f(\lambda_\parallel,\lambda_\perp, \tau)$ for case I (upper picture)
and for case II  (bottom picture)  for different values of $\lambda_\parallel$ and $\lambda_\perp$ and for fixed overall perturbation $\lambda$ $\equiv$ $\sqrt{\lambda_\parallel^2+\lambda_\perp^2}$ $ = 0.1$.  The values of $\lambda_\parallel$ and $\lambda_\perp$ are given by the  five ratios  $\lambda_\parallel^2/\lambda_\perp^2$ $=$ $\infty$ (full line), $2$ (dotted line), $1$ (dashed line), $1/2$ (dashed--dotted line), $0$ (dashed--dotted-dotted--dotted line).}
  \label{fig:fid2}
\end{figure}

 In figure \ref{fig:fid2} fidelity amplitude is plotted for small perturbation strength $\lambda=0.1$ and
for different ratios between $\lambda_\parallel$ and $\lambda_\perp$ for case I and case II.

In case I we see that fidelity amplitude is a monotonous function 
of this ratio for all times. The slowest decay happens for $\lambda_\parallel =0$, i.~e. 
when the perturbation in the direction of $H_0$ is zero (freeze). In
case II fidelity shows qualitatively the same behavior, i.~e.~ a slower decay for perpendicular perturbations for times beyond Heisenberg time. 
This suggest to define for a general perturbation $V$ 
\begin{equation}
\tr H_0 V \ = \ 0
\end{equation}
 as  condition for a fidelity freeze, which is slightly more general than the one proposed in \cite{pro05}.  However in case II the freeze is much less pronounced than in case I, indicating that the diagonal elements of  $V_\perp$, albeit $\tr V_\perp=\tr V_\perp H_0 = 0$, have some impact on the decay. 
 
 A careful look reveals that for times beyond Fermi's golden rule but smaller than Heisenberg time in case II fidelity decay is slower for a parallel perturbations than for a perpendicular perturbation. 
 
\begin{figure}
\includegraphics[width=1\linewidth,height=5cm]{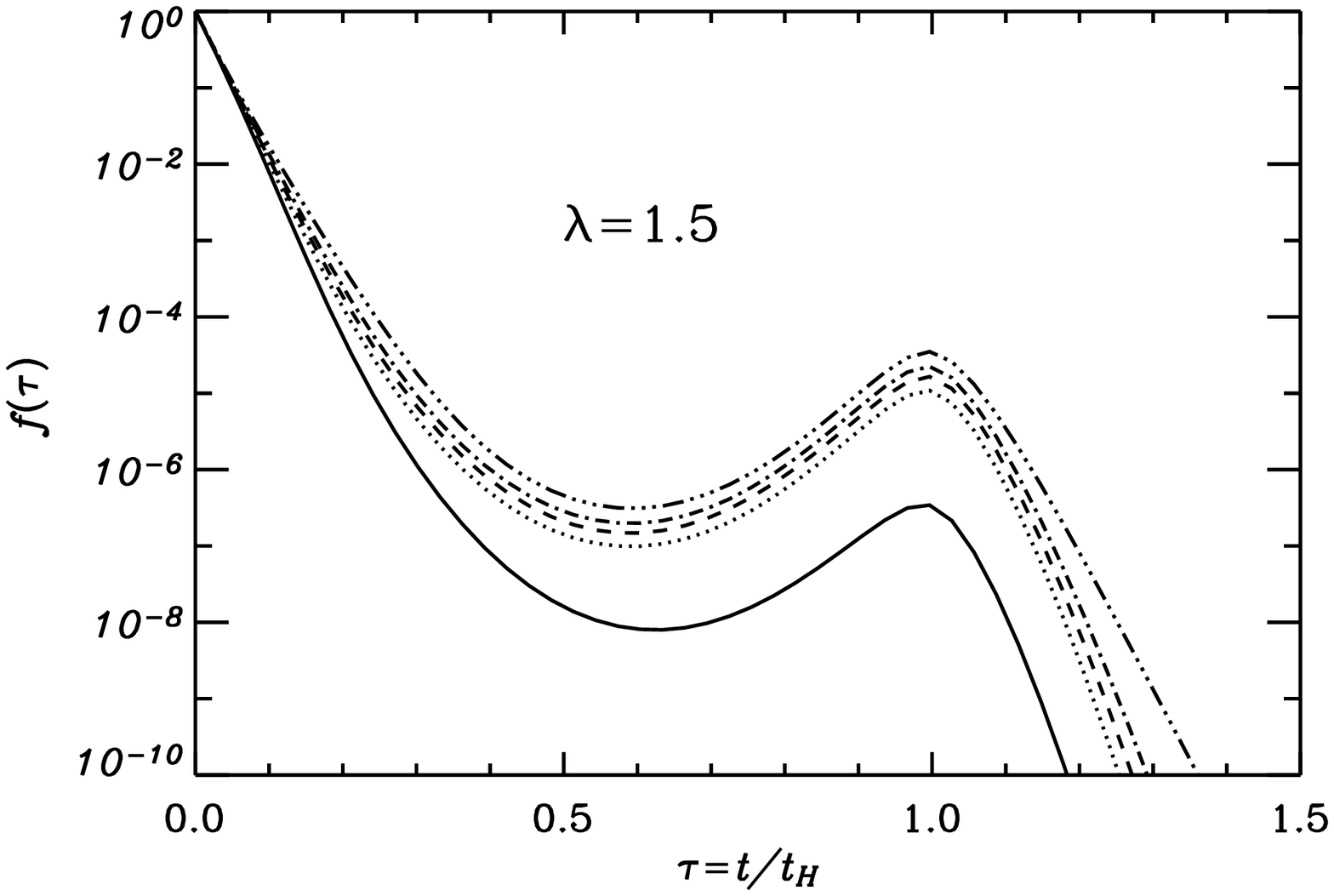}\\
\includegraphics[width=1\linewidth,height=5cm]{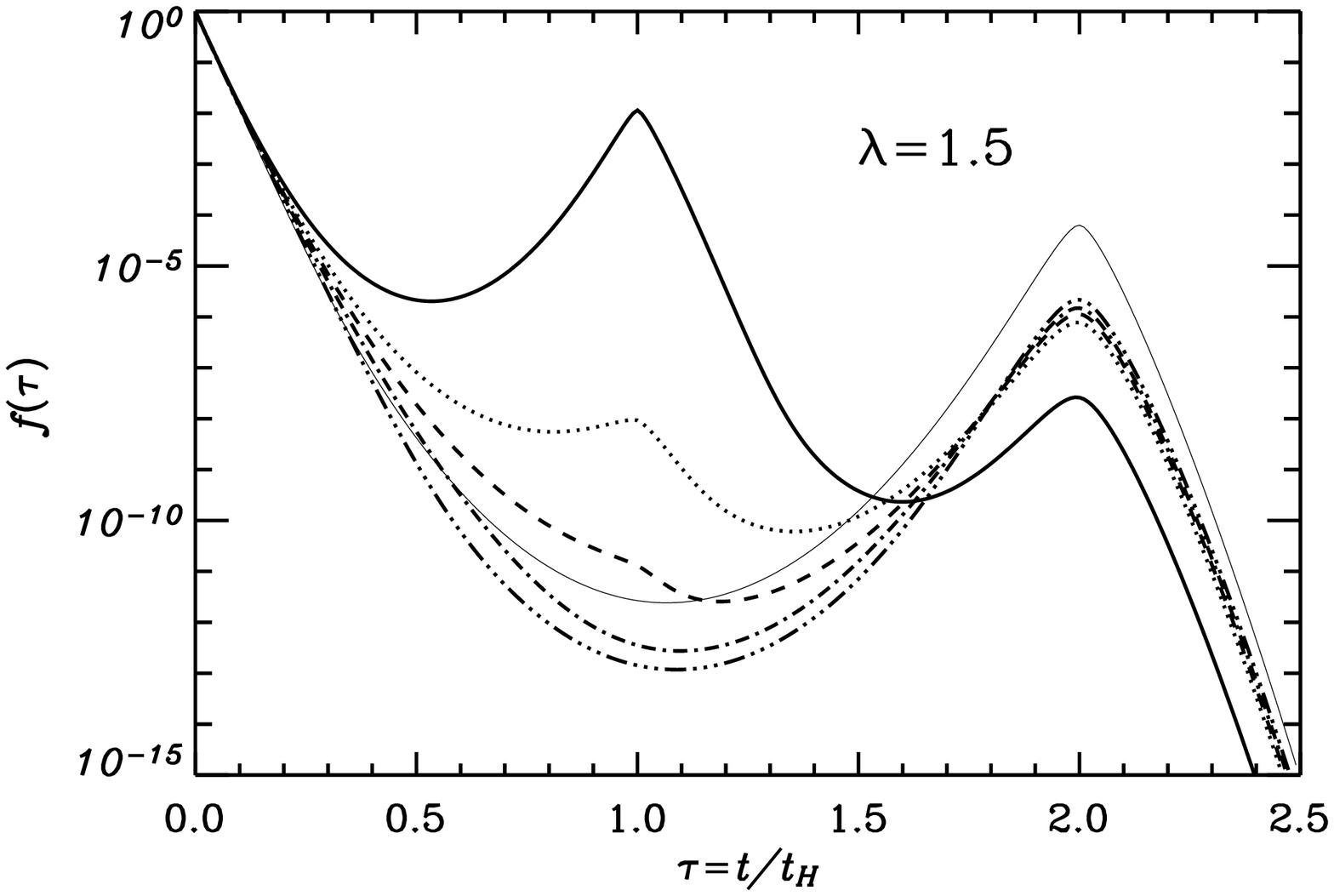}\\
  \caption{ Same as figure \ref{fig:fid2} but for a strong overall perturbation $\lambda=1.5$.  
Fidelity amplitude $f(\lambda_\parallel,\lambda_\perp, \tau)$ is shown for case I (upper panel) and case II (lower panel) and for the ratios  $\lambda_\parallel^2/\lambda_\perp^2$ $=$ $\infty$ (full line), $2$ (dotted line), $1$ (dashed line), $1/2$ (dashed--dotted line), $0$ (dashed--dotted-dotted--dotted line).
In the lower panel  for comparison fidelity amplitude $f(\sqrt{2}\lambda, \tau/2)$ for a GUE with a GUE perturbation is shown as well (thinner full line).}
  \label{fig:fid2a}
\end{figure}

This becomes evident for strong perturbations. In figure \ref{fig:fid2a} fidelity amplitude is plotted for the same ratios of $\lambda_\parallel$ and $\lambda_\perp$ as before but for strong overall perturbation $\lambda=1.5$. Case I fidelity decay shows monotonous behavior as a function $\lambda_\parallel/\lambda_\perp$ and fidelity decay is for all times slowest for a perpendicular perturbation. 
However case II fidelity decay is more complicated.  For times smaller than Heisenberg time decay is slowest for a purely parallel perturbation and fastest for a purely perpendicular one. At Heisenberg time a pronounced revival is seen for a purely parallel perturbation. The peak decreases as the share of the perpendicular perturbation increases. Finally for a purely perpendicular perturbation there is a minimum at Heisenberg time and no revival at all. 

After Heisenberg time things change. Now decay becomes fastest for a purely parallel perturbation with only a tiny second revival at  twice the Heisenberg time. For a purely perpendicular perturbation the freeze behavior comes in and at twice the Heisenberg time a sizable revival occurs, such that just as in case I for long times decay is slowest for a purely perpendicular perturbation.  Somewhere between Heisenberg time and twice the Heisenberg time the two curves cross.   

To understand  this behavior qualitatively, we recall two peculiarities of the GSE: first the spectral rigidity is much higher than for the GUE or the GOE. It has been argued \cite{sto05} that the revival at Heisenberg time is a signature of the high spectral rigidity. More generally high spectral rigidity favors a slow decay. Second the eigenvalues of the GSE are two--fold degenerate (Kramers degeneracy).

Thus a perpendicular perturbation has two effects: first it breaks time reversal invariance and drives the GSE into a GUE. Since the latter has lower spectral rigidity, this has as consequence that the peak at Heisenberg time becomes less and lesser pronounced and for times smaller than Heisenberg time decay is enhanced by the perpendicular perturbation. Second it breaks Kramers degeneracy, thus the number of independent levels and therefore level density and Heisenberg time double. This leads to the pronounced peak at twice the (original) Heisenberg time.  A comparison with the plot of fidelity amplitude $f(\sqrt{2}\lambda,\tau/2)$ of a GUE, with a GUE perturbation(${\rm A}+\lambda{\rm A}$) shows indeed good agreement.

\begin{figure}
\includegraphics[width=1\linewidth,height=5cm]{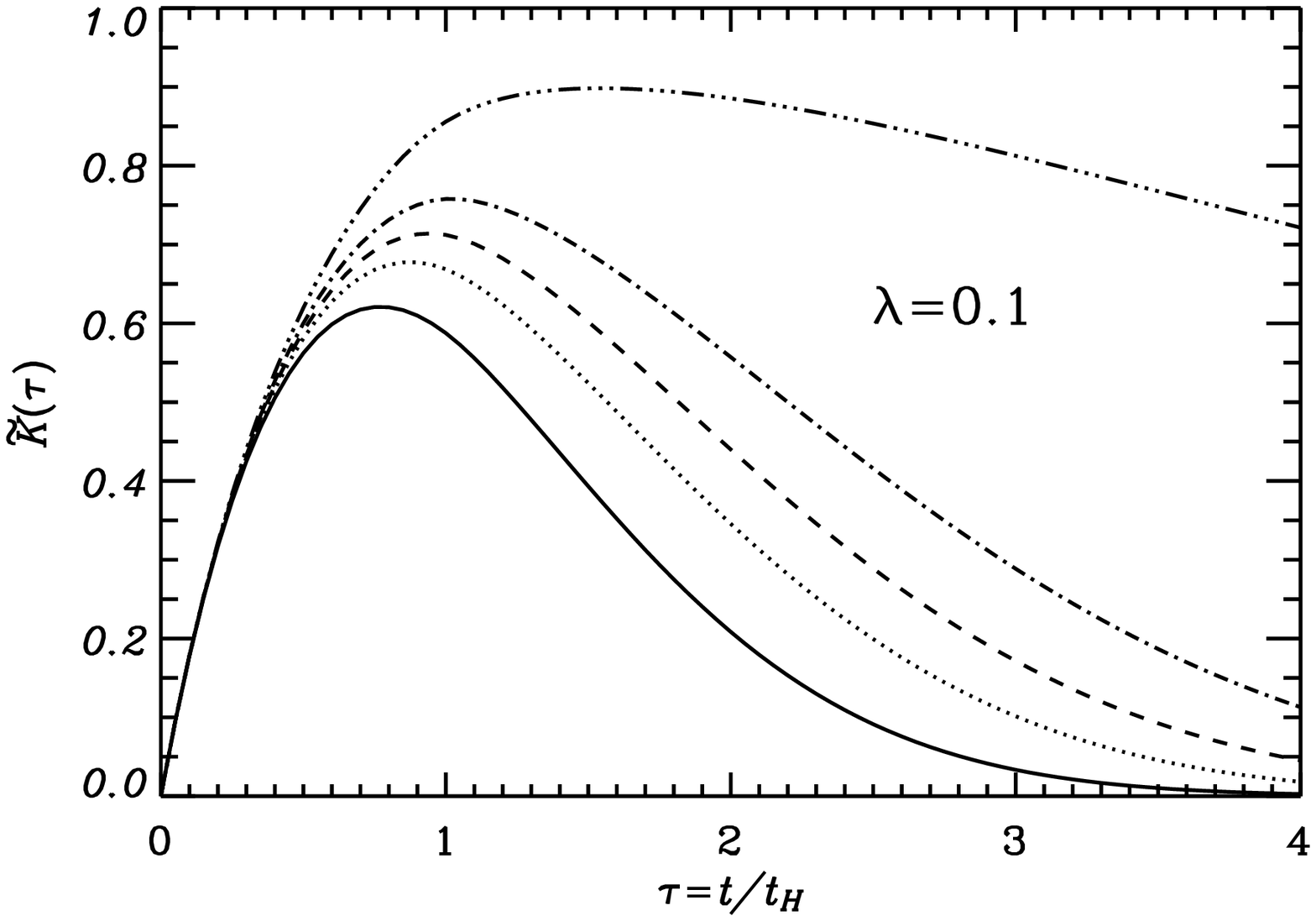}\hfill
\includegraphics[width=1\linewidth,height=5cm]{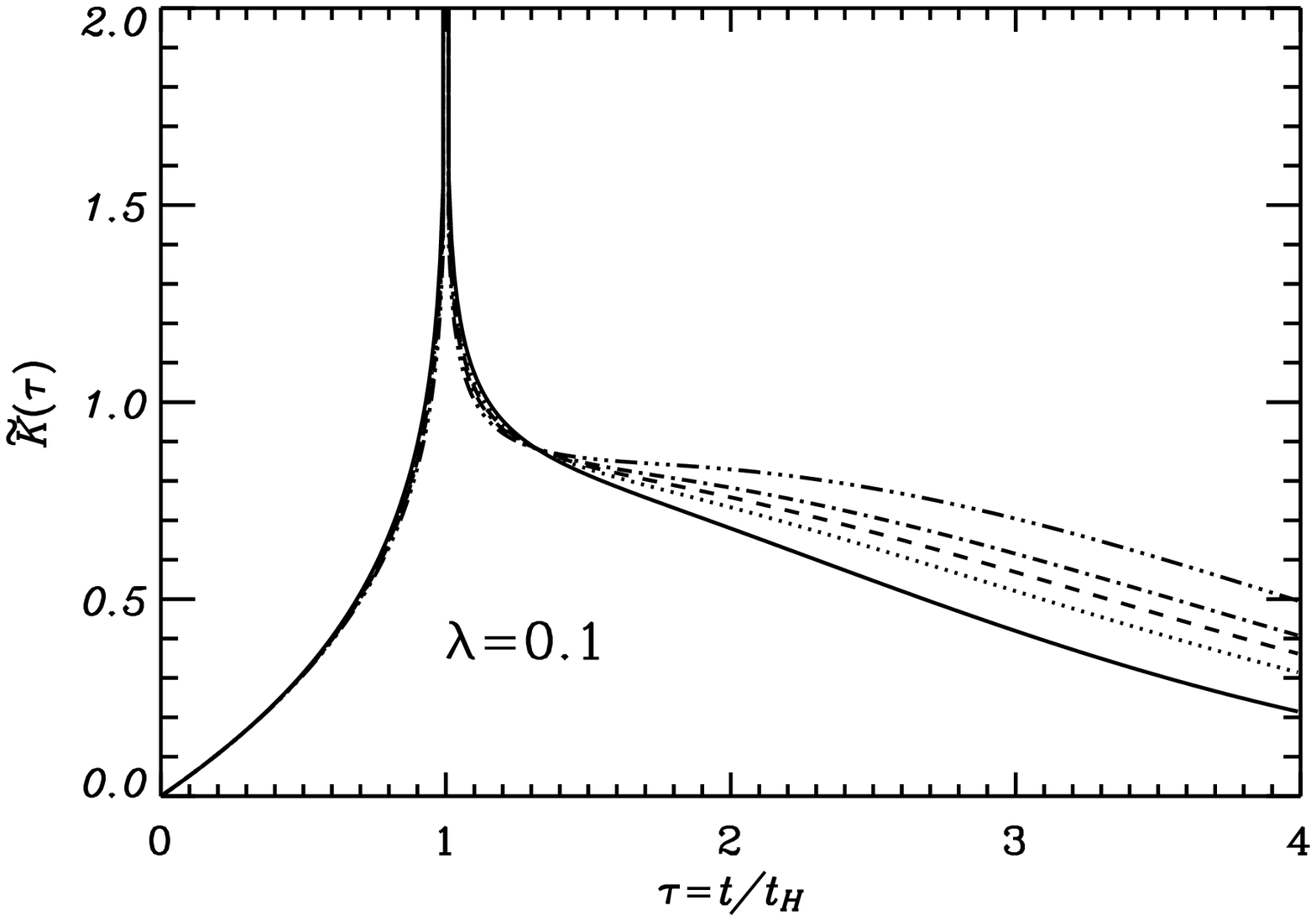}\\
  \caption{Cross form--factor $\tilde{K}(\lambda_\parallel,\lambda_\perp, \tau)$  for case I (upper panel) and case II (lower panel)
 for different values of $\lambda_\parallel$ and $\lambda_\perp$ and for small overall 
 perturbation $\lambda=0.1$. The values of  $\lambda_\parallel$ and $\lambda_\perp$ are given by the the ratios
$\lambda_\parallel^2/\lambda_\perp^2$ $=\infty$ (full line), $2$ (dotted line), $1$ (dashed line), $1/2$ (dashed--dotted line), $0$ (dashed--dotted-dotted--dotted line).}
  \label{fig:fid3}
\end{figure}

In figure \ref{fig:fid3} the cross form factor is plotted in both cases for the same five ratios between $\lambda_\parallel$ and $\lambda_\perp$ as before. Qualitatively the behavior is similar to fidelity amplitude. In case I the form factor is smallest for a purely parallel perturbation for all times. In case II before Heisenberg time the form factor is smallest for a purely perpendicular perturbation and largest for a purely parallel one. After Heisenberg time the order is inverted. At Heisenberg time a logarithmic singularity occurs, which is typical for the GSE. For strong perturbations the cross form--factor develops peaks at Heisenberg time and for case II at twice the Heisenberg time (not shown here). It has its cause in the algebraic decay of the cross form--factor  at these specific times \cite{koh08}. At all other times it decays exponentially.

\section{Conclusion}
In conclusion we presented the analytic formulae for fidelity amplitude and cross--form factor for parametric RMT ensembles, where the 
time reversal invariance of the unperturbed system is broken by the perturbation. The general perturbation is split into a parallel component, sharing the symmetries of the original Hamiltonian and a perpendicular component which maximally breaks this symmetry.  

Both possibilities of TRS breaking, even spin GOE$\to$ GUE and odd spin GSE$\to$ GUE, were discussed on equal footing. In the first case a strong freeze effect occurs for a purely perpendicular perturbation. It can be explained by the absence of diagonal elements of the perturbation in the eigenbasis of the unperturbed Hamiltonian. In case II long time decay is slowest for a purely perpendicular perturbation as well. This leads us to propose $\tr H_0V=0$ as a more general condition for a reduced fidelity decay.  However  in case II the perturbation has diagonal entries in the eigenbasis of $H_0$ and the attenuation of decay is much less pronounced than in case I.  We are reluctant to call this behavior "freeze". We propose to call it "weak fidelity freeze". 

The full Hilbert space is involved in the condition $\tr H_0V=0$. Therefore it is only applicable to fidelity decay with respect to a random initial state as considered here, to which all states of the Hilbert space contribute.  For an arbitrary initial state this condition will in general not suffice to attenuate fidelity decay.

In the differential relation between fidelity and cross--form factor only the parallel perturbation strength enters. The relation 
might be verified experimentally for instance in a billiard experiment as described in \cite
{die09}.  It might be  used to measure fidelity indirectly via spectral correlations.
 
\begin{acknowledgments}
The authors acknowledge financial support from the Deutsche Forschungsgemeinschaft (DFG) 
with grant No.~KO3538/1-2 (HK) and via the 
research group 760 "Scattering systems with complex dynamics" (HJS) and from the Japan Society 
for the Promotion 
of Science (KAKENHI 20540372) (TN). TN thanks Dr. Keiji 
Saito for valuable discussions. TN also thanks the University of Duisburg--Essen for its hospitality.
\end{acknowledgments}
 
\appendix

\section{Derivation of Eq.~(\ref{mres4}) based on a universality argument}
\label{appB}

In this section we demonstrate on the example of the transition GOE $\to$ GUE (case I) 
 how the method of Refs.~\cite{koh08,smo08} can be extended to 
the case of symmetry breaking.  We introduce new variables
\begin{equation}
{\tilde \lambda}_\parallel = \frac{\lambda_\parallel}{2}, \ \  
{\tilde H}_0 = H_0 + \frac{\lambda_\parallel}{2} V_\parallel \ .
\end{equation}
For  ${\tilde H}_0$  we allow for a general probability measure in the GOE universality class and denote it by $d\nu({\tilde H}_0)$, while those of $V_\parallel$ and $V_\perp$ 
are Gaussian measures as before ($d V_\parallel$ and $d V_\perp$ include the 
normalization constants). Since the probability measure of ${\tilde H}_0$ 
is assumed to be general, it should be typical and free from any special 
constraint besides the matrix symmetry. 
\par
Now we define    
\begin{eqnarray}
{\tilde {\cal F}}_{\alpha \beta,\gamma \delta} & = &  
\left( \frac{1}{z_1 - H_0} \right)_{\alpha \beta} \
\left( \frac{1}{z_2 - H} 
\right)_{\gamma \delta}
\nonumber \\ & = &  
\left( \frac{1}{z_1 - {\tilde H}_0 + {\tilde \lambda}_\parallel V_\parallel} \right)_{\alpha \beta} \nonumber\\
&&\quad \times\left( \frac{1}{z_2 - {\tilde H}_0 - {\tilde \lambda}_\parallel V_\parallel - i \lambda_\perp V_\perp} 
\right)_{\gamma \delta} \ .
\end{eqnarray}
Introducing delta--distributions of matrix arguments we can express ${\tilde {\cal F}}_{\alpha \beta,\gamma \delta} $
as
\begin{eqnarray}
{\tilde {\cal F}}_{\alpha \beta,\gamma \delta} 
& = & \int dH_1 d H_2 \ \delta(H_1 - {\tilde H}_0 
+ {\tilde \lambda}_\parallel V_\parallel) {\cal F}_{\alpha \beta,\gamma\delta}  \\ 
& \times &  
\delta(H_2^{(R)} - {\tilde H}_0 - {\tilde \lambda}_\parallel V_\parallel) \ 
\delta(H_2^{(I)} - \lambda_\perp V_\perp) \ \nonumber
\end{eqnarray}
where $H_1$ is an $N \times N$ real symmetric matrix, 
$H_2$ is an $N \times N$ hermitian matrix and $H_2^{(R)}= {\rm Re}H_2$, $H_2^{(I)}= {\rm Im}H_2$. Moreover
\begin{equation}
{\cal F}_{\alpha \beta,\gamma \delta} = 
\left( \frac{1}{z_1 - H_1} \right)_{\alpha \beta} 
\left( \frac{1}{z_2 - H_2} \right)_{\gamma \delta} \ .
\end{equation}
All three delta--distributions can be Fourier transformed. We find
\begin{eqnarray}
{\tilde {\cal F}}_{\alpha \beta,\gamma \delta} & = & \int d\Lambda_1 d\Lambda_2 d\Lambda_3 
d H_1 d H_2 \ 
e^{2 \pi i {\rm tr} \Lambda_1 (H_1 - {\tilde H}_0 + 
{\tilde \lambda}_\parallel V_\parallel)} \nonumber \\ 
& &\qquad \times 
e^{2 \pi i {\rm tr} \Lambda_2 
(H_2^{(R)} - {\tilde H}_0 - {\tilde \lambda}_\parallel V_\parallel)} \nonumber\\ 
&&\qquad \times e^{2 \pi i {\rm tr} \Lambda_3 (H_2^{(I)} - 
\lambda_\perp V_\perp)} \ {\cal F}_{\alpha \beta,\gamma \delta}\quad .
\end{eqnarray}
Here $\Lambda_{1,2,3}$ are matrices which have the same symmetry as their real space counterparts, namely $H_1$, $H_2^{(R)}$ and $H_2^{(I)}$. 
This means $\Lambda_1$  and $ \Lambda_2$ are  $N \times N$  real symmetric matrices and    
$\Lambda_3$ is an  $N \times N $ real antisymmetric matrix. The integration domain is the real axis for all 
independent entries of $\Lambda_{n}$, $n=1,2,3$. 
The expectation value of ${\tilde {\cal F}}_{\alpha \beta, \gamma \delta}$ 
can be written as
\begin{eqnarray}
\label{ref1}
\left\langle {\cal F}_{\alpha \beta, \gamma \delta} \right\rangle& =& 
\int d\nu({\tilde H}_0) dV_\parallel dV_\perp  {\tilde {\cal F}}_{\alpha \beta,\gamma \delta} \nonumber\\
  &&\quad\times e^{- (1/4){\rm tr}V_\parallel^2 
+ (1/4){\rm tr}V_\perp^2}
\nonumber \\ 
& = & \int d\nu({\tilde H}_0) d\Lambda_1 d\Lambda_2 d\Lambda_3 
dH_1 dH_2 {\cal F}_{\alpha \beta,\gamma \delta} \nonumber\\
&&\times  e^{- (2 \pi {\tilde \lambda}_\parallel)^2 
{\rm tr} (\Lambda_1 - \Lambda_2)^2 + (2 \pi \lambda_\perp)^2 
{\rm tr} (\Lambda_3)^2} \label{37} \\ 
& &\times  e^{2 \pi i {\rm tr}\left\{ \Lambda_1 (H_1 - {\tilde H}_0) 
+ \Lambda_2 (H_2^{(R)} - {\tilde H}_0) + \Lambda_3 H_2^{(I)} \right\} } . \nonumber
\end{eqnarray}
Here the brackets $\langle \ldots\rangle$ do not simply mean the expectation 
value. Rather $\langle {\cal F}_{\alpha \beta,\gamma \delta} \rangle$ is defined to 
be the expectation value of ${\tilde {\cal F}}_{\alpha \beta,\gamma \delta}$.

Now we introduce the notation
\begin{eqnarray}
{\rm tr}\frac{\partial^2}{\partial H_1 \partial H_2^{(R)}} 
&=& \sum_{j=1}^N \frac{\partial^2}{\partial 
(H_1)_{jj} \ \partial \left( H_2^{(R)} \right)_{jj}} \nonumber\\   
&&+ \frac{1}{2}\sum_{j<l}^N \frac{\partial^2}{\partial 
(H_1)_{jl} \ \partial \left( H_2^{(R)} \right)_{jl}}.
\end{eqnarray}  
Then it follows from partial integrations that
\begin{equation}
\label{39}
\left\langle {\rm tr} \frac{\partial^2}{\partial H_1 
\partial H_2^{(R)}} 
{\cal F}_{\alpha \alpha, \beta \beta} \right\rangle = - (2 \pi)^2 
\left\langle {\rm tr}(\Lambda_1 \Lambda_2) {\cal F}_{\alpha \alpha,
\beta \beta} \right\rangle  
\end{equation}
and
\begin{equation}
\label{40}
\frac{\partial}{\partial ({\tilde \lambda}_\parallel^2)} \left\langle 
{\cal F}_{\alpha \alpha, \beta \beta} \right\rangle = - (2 \pi)^2 
\left\langle {\rm tr}(\Lambda_1 - \Lambda_2)^2 {\cal F}_{\alpha \alpha,
\beta \beta} \right\rangle.  
\end{equation}
Here repeated indices stand for summations from $1$ to $N$.

Let us note that a simultaneous shift of $H_1$ and $H_2^{(R)}$ in (\ref{ref1}) 
induces a shift of ${\tilde H}_0$. Although such a shift modifies the 
measure $d\nu({\tilde H}_0)$, the universality of the spectral 
correlation function implies that $\langle {\cal F}_{\alpha \alpha,\beta 
\beta} \rangle$ is asymptotically invariant in the limit $N \rightarrow 
\infty$.  Therefore we obtain the following estimate
\begin{eqnarray}
&&\left\langle {\rm tr}\left( 
\frac{\partial}{\partial H_1} + 
\frac{\partial}{\partial H_2^{(R)}} \right)^2 
{\cal F}_{\alpha \alpha,\beta \beta} \right\rangle 
= \nonumber\\
&&\qquad\qquad- (2 \pi)^2 \langle {\rm tr}(\Lambda_1 + \Lambda_2)^2 
{\cal F}_{\alpha \alpha,\beta \beta} \rangle \approx 0.
\label{estimate}
\end{eqnarray}
From this it follows that
\begin{eqnarray}
\frac{\partial}{\partial (\lambda_\parallel^2)} 
\langle {\cal F}_{\alpha \alpha,\beta \beta} 
\rangle 
& + & \left\langle {\rm tr}\frac{\partial^2}{
\partial H_1 \partial H_2^{(R)}} 
{\cal F}_{\alpha \alpha, \beta \beta} 
\right\rangle \nonumber \\ 
& & = - \pi^2 \langle {\rm tr}(\Lambda_1 + \Lambda_2)^2 
{\cal F}_{\alpha \alpha,\beta \beta} \rangle 
\nonumber \\ 
& & \approx 0. \label{ref2}
\end{eqnarray}
\par
In order to show that the estimate (\ref{estimate}) is indeed 
correct, let us pay attention to Eq.~(\ref{ref1}). Proper 
unfolding of the energy level correlations requires 
an ${\cal O}(1)$ scaling of the eigenvalue density of ${\tilde H}_0$. 
Each element of the perturbation $V_\parallel$ is set to be ${\cal O}(1)$, 
because it should equally scale as the mean level spacing. 
When the eigenvalue density is scaled 
as ${\cal O}(1)$, since there are $N$ eigenvalues, each 
eigenvalue ${\tilde E}_{0j}$ of ${\tilde H}_0$ should 
typically be ${\cal O}(N)$. Then the RHS of the identity 
\begin{equation}
{\rm tr}({\tilde H}_0)^2 = \sum_{j=1}^N ({\tilde E}_{0j})^2
\end{equation}
becomes ${\cal O}(N^3)$. In the LHS, on the other hand, we 
have ${\cal O}(N^2)$ terms, each of which is the square 
of an element of ${\tilde H}_0$. Therefore each element of ${\tilde H_0}$ 
is estimated as ${\cal O}(N^{1/2})$. Then the main contribution 
to the integral over the matrix $\tilde{H}_0$ with respect to 
the measure $d\nu(\tilde{H}_0)$ in equation (\ref{ref1}) 
comes from a region where the elements of $\Lambda_1 + \Lambda_2$ 
are of order ${\cal O}(N^{-1/2})$. Only in that region a rapid 
oscillation of the exponential factor is avoided. 

It can be seen from the Gaussian factor in Eq.(\ref{ref1})
that the elements of $\Lambda_1 - \Lambda_2$
are scaled as ${\cal O}(1)$. Because of the identity
\begin{equation}
(\Lambda_1 - \Lambda_2)^2 = - 2 (\Lambda_1 \Lambda_2 
+ \Lambda_2 \Lambda_1) + (\Lambda_1 + \Lambda_2)^2, 
\end{equation}
the elements of $(\Lambda_1 - \Lambda_2)^2$ are approximated 
by the elements of $- 2 (\Lambda_1 \Lambda_2 + \Lambda_2 
\Lambda_1)$. Hence we find an estimate
\begin{equation}
{\rm tr}(\Lambda_1 - \Lambda_2)^2 \approx 
- 4 {\rm tr}(\Lambda_1 \Lambda_2),
\end{equation}
which implies Eq.(\ref{ref2}). We notice that this estimate can only 
be fulfilled when ${\rm tr}(\Lambda_1 \Lambda_2)$ is negative.

On the other hand, we can readily find 
\begin{equation}
{\rm tr}\frac{\partial^2}{\partial H_1 \partial H_2^{(R)}}
{\cal F}_{\alpha \alpha,\beta \beta} \ = \
{\rm tr} \left( \frac{1}{z_1 - H_1} \right)^2 
\left( \frac{1}{z_2 - H_2} \right)^2 
\end{equation}
and
\begin{eqnarray}
\frac{\partial^2}{\partial z_1 \partial z_2} 
{\cal F}_{\alpha \beta,\beta \alpha} & = &  
\frac{\partial^2}{\partial z_1 \partial z_2}   
\left( \frac{1}{z_1 - H_1} \right)_{\alpha \beta} 
\left( \frac{1}{z_2 - H_2} \right)_{\beta \alpha} 
\nonumber \\ 
& = &  
{\rm tr} \left( \frac{1}{z_1 - H_1} \right)^2  
\left( \frac{1}{z_2 - H_2} \right)^2, 
\end{eqnarray}
so that
\begin{equation}
\label{ref3}
\left\langle {\rm tr}\frac{\partial^2}{\partial H_1 \partial H_2^{(R)}} 
{\cal F}_{\alpha \alpha,\beta \beta}  
\right\rangle = 
\frac{\partial^2}{\partial z_1 \partial z_2}  
\left\langle
{\cal F}_{\alpha \beta,\beta \alpha} \right\rangle.  
\end{equation}
Comparing (\ref{ref2}) and (\ref{ref3}), we arrive at
\begin{equation}
\frac{\partial}{\partial (\lambda_\parallel^2)} \left\langle 
{\cal F}_{\alpha \alpha, \beta \beta} \right\rangle 
\approx - \frac{\partial^2}{\partial z_1 \partial z_2}  
\left\langle {\cal F}_{\alpha \beta,\beta \alpha} \right\rangle,  
\end{equation}
which gives the required relation (\ref{mres4}) between the fidelity and 
parametric spectral correlation, respectively cross form--factor.

\section{Derivation of equations (\ref{mres1}) and (\ref{mres4}) from Ref.~\cite{tan94}}
\label{appA}

In Ref.~\cite{tan94}, called THSA in the following, the Fourier--transform of the cross form--factor 
was derived as a three--fold integral
\begin{eqnarray}
K(\bar{x},x_o,x_u,\omega) &=& {\rm Re}\int d\lambda d\lambda_1d\lambda_2W e^{F_\pm} \ ,
\end{eqnarray}
where the integration domains  are in case I defined by $\lambda\in [-1,1]$, $\lambda_1\in 
[1,\infty]$, $\lambda_2\in [1,\infty]$ and in case II by $\lambda\in [1,\infty]$, 
$\lambda_1\in [-1,1]$, $\lambda_2\in [0,1]$.
Setting the parameter $\bar{x} = x_u/2$ the expressions for $F$ and $W$ (equations (5) and (6) of THSA) are given by
\begin{eqnarray}
F_\pm& = &\pm\kappa i\pi\omega(\lambda_1\lambda_2-\lambda) \pm \frac{x_u^2\pi^2}{2}\left(\lambda_1^2+
\lambda_2^2-\lambda^2-1\right)\nonumber\\
    &&       \pm \frac{x_o^2\pi^2}{4}\left(2\lambda_1^2\lambda_2^2-\lambda^2-\lambda_1^2-
\lambda_2^2+1\right)\\
W&=& \frac{\left(\lambda_1\lambda_2-\lambda\right)^2\left(1-\lambda^2\right)}{\left(\lambda_1^2+\lambda_2^2+\lambda^2-2\lambda_1\lambda_2-1\right)^2}\times\nonumber\\ \label{Wfunc}
&&\left(1+\frac{\pi^2x_u^2}{\kappa}\left(\lambda_1^2+\lambda_2^2+\lambda^2-2\lambda\lambda_1\lambda_2-1\right)\right)\ .
\end{eqnarray}
Here the plus sign applies to case I and the minus sign to case II. The parameter $\kappa$ has the value $\kappa=1$ (case I) and $\kappa=2$ (case II). 
This factor does not appear in THSA, however it does appear in Ref.~\cite{sim93a}. We introduced it, such that $\tilde{K}(t)$ is related to
$K(\omega)$ in both cases via 
\begin{equation}
\tilde{K}(\tau) \ = \ \int d\omega e^{-2\pi i \omega \tau} K(\omega) \ .
\end{equation}
In THSA the function $W$ differs in case I and case II by a relative minus sign between two summands in the last line of equation (\ref{Wfunc}) . This seems to be wrong. Moreover in the same line the factor $1/\kappa$ in the second summand is missing in THSA. 

Fourier transformation yields $\delta(\tau-\lambda_1\lambda_2/2+\lambda)$ in case I and  $\delta(\tau-\lambda+\lambda_1\lambda_2)$ in case II, which allows to integrate over $\lambda$. Equations (\ref{mres1}) to (\ref{mres4}) are obtained through the transformations
\begin{eqnarray}
&& \left.\begin{array}{ccl}
        u&=& \frac{1}{2}\left(\lambda_1\lambda_2-1\right)\\[0.3em]
v&=& \frac{1}{2}\sqrt{\lambda_1^2\lambda_2^2-\lambda_1-\lambda_2+1}\end{array} \right\}\mbox{\rm case I}\\
&& \left.\begin{array}{ccl}
        u&=& \lambda_1\lambda_2\\[0.3em]
v&=& \sqrt{\lambda_1^2\lambda_2^2-\lambda_1-\lambda_2+1}\end{array} \right\}\mbox{\rm case II.}
\end{eqnarray}
The parameters are identified as $\lambda_\parallel = x_o/2$ and $\lambda_\perp = x_u/\sqrt{2}$.

\begin{thebibliography}{24}
\expandafter\ifx\csname natexlab\endcsname\relax\def\natexlab#1{#1}\fi
\expandafter\ifx\csname bibnamefont\endcsname\relax
  \def\bibnamefont#1{#1}\fi
\expandafter\ifx\csname bibfnamefont\endcsname\relax
  \def\bibfnamefont#1{#1}\fi
\expandafter\ifx\csname citenamefont\endcsname\relax
  \def\citenamefont#1{#1}\fi
\expandafter\ifx\csname url\endcsname\relax
  \def\url#1{\texttt{#1}}\fi
\expandafter\ifx\csname urlprefix\endcsname\relax\def\urlprefix{URL }\fi
\providecommand{\bibinfo}[2]{#2}
\providecommand{\eprint}[2][]{\url{#2}}

\bibitem[{\citenamefont{Gorin et~al.}(2006)\citenamefont{Gorin, Prosen,
  Seligman, and \v{Z}nidari\v{c}}}]{gor06}
\bibinfo{author}{\bibfnamefont{T.}~\bibnamefont{Gorin}},
  \bibinfo{author}{\bibfnamefont{T.}~\bibnamefont{Prosen}},
  \bibinfo{author}{\bibfnamefont{T.~H.} \bibnamefont{Seligman}},
  \bibnamefont{and}
  \bibinfo{author}{\bibfnamefont{M.}~\bibnamefont{\v{Z}nidari\v{c}}},
  \bibinfo{journal}{Phys. Rep.} \textbf{\bibinfo{volume}{435}},
  \bibinfo{pages}{33} (\bibinfo{year}{2006}).

\bibitem[{\citenamefont{Jacquod and Petitjean}(2009)}]{jac09}
\bibinfo{author}{\bibfnamefont{P.}~\bibnamefont{Jacquod}} \bibnamefont{and}
  \bibinfo{author}{\bibfnamefont{C.}~\bibnamefont{Petitjean}},
  \bibinfo{journal}{Adv. Phys.} \textbf{\bibinfo{volume}{58}},
  \bibinfo{pages}{67} (\bibinfo{year}{2009}).

\bibitem[{\citenamefont{Sch{\"a}fer et~al.}(2005)\citenamefont{Sch{\"a}fer,
  Gorin, Seligman, and St{\"o}ckmann}}]{schae05b}
\bibinfo{author}{\bibfnamefont{R.}~\bibnamefont{Sch{\"a}fer}},
  \bibinfo{author}{\bibfnamefont{T.}~\bibnamefont{Gorin}},
  \bibinfo{author}{\bibfnamefont{T.~H.} \bibnamefont{Seligman}},
  \bibnamefont{and} \bibinfo{author}{\bibfnamefont{H.~J.}
  \bibnamefont{St{\"o}ckmann}}, \bibinfo{journal}{New J. Phys.}
  \textbf{\bibinfo{volume}{7}}, \bibinfo{pages}{152} (\bibinfo{year}{2005}).

\bibitem[{\citenamefont{H{\"o}hmann et~al.}(2008)\citenamefont{H{\"o}hmann,
  Kuhl, and St{\"o}ckmann}}]{hoe08}
\bibinfo{author}{\bibfnamefont{R.}~\bibnamefont{H{\"o}hmann}},
  \bibinfo{author}{\bibfnamefont{U.}~\bibnamefont{Kuhl}}, \bibnamefont{and}
  \bibinfo{author}{\bibfnamefont{H.-J.} \bibnamefont{St{\"o}ckmann}},
  \bibinfo{journal}{Phys. Rev. Lett.} \textbf{\bibinfo{volume}{100}},
  \bibinfo{pages}{124101} (\bibinfo{year}{2008}).

\bibitem[{gen()}]{genpar}
\bibinfo{note}{B. D. Simons, P. A. Lee, and B. L. Altshuler, Phys. Rev. Lett.
  {\bf 72}, 64 (1994); B. D. Simons {\it et al.}, Phys. Rev. Lett. {\bf 71},
  2899 (1993); M. Faas {\it et al.}, Phys. Rev. B {\bf 48}, 5439 (1993);
  B.~Dietz, M. Lombardi and T.~H.~Seligman, Phys. Lett. A {\bf 215}, 181
  (1996).}

\bibitem[{\citenamefont{Bertelsen et~al.}(1999)\citenamefont{Bertelsen,
  Ellegaard, Guhr, Oxborrow, and Schaadt}}]{ber99}
\bibinfo{author}{\bibfnamefont{P.}~\bibnamefont{Bertelsen}},
  \bibinfo{author}{\bibfnamefont{C.}~\bibnamefont{Ellegaard}},
  \bibinfo{author}{\bibfnamefont{T.}~\bibnamefont{Guhr}},
  \bibinfo{author}{\bibfnamefont{M.}~\bibnamefont{Oxborrow}}, \bibnamefont{and}
  \bibinfo{author}{\bibfnamefont{K.}~\bibnamefont{Schaadt}},
  \bibinfo{journal}{Phys. Rev. Lett.} \textbf{\bibinfo{volume}{83}},
  \bibinfo{pages}{2171} (\bibinfo{year}{1999}).

\bibitem[{\citenamefont{Hul et~al.}(2009)\citenamefont{Hul, \v{S}eba, and
  Sirko}}]{hul09}
\bibinfo{author}{\bibfnamefont{O.}~\bibnamefont{Hul}},
  \bibinfo{author}{\bibfnamefont{P.}~\bibnamefont{\v{S}eba}}, \bibnamefont{and}
  \bibinfo{author}{\bibfnamefont{L.}~\bibnamefont{Sirko}},
  \bibinfo{journal}{Physica Scripta} \textbf{\bibinfo{volume}{2009}},
  \bibinfo{pages}{014048} (\bibinfo{year}{2009}).

\bibitem[{\citenamefont{Kohler et~al.}(2008)\citenamefont{Kohler, Smolyarenko,
  Pineda, Guhr, Leyvraz, and Seligman}}]{koh08}
\bibinfo{author}{\bibfnamefont{H.}~\bibnamefont{Kohler}},
  \bibinfo{author}{\bibfnamefont{I.~E.} \bibnamefont{Smolyarenko}},
  \bibinfo{author}{\bibfnamefont{C.}~\bibnamefont{Pineda}},
  \bibinfo{author}{\bibfnamefont{T.}~\bibnamefont{Guhr}},
  \bibinfo{author}{\bibfnamefont{F.}~\bibnamefont{Leyvraz}}, \bibnamefont{and}
  \bibinfo{author}{\bibfnamefont{T.~H.} \bibnamefont{Seligman}},
  \bibinfo{journal}{Phys. Rev. Lett.} \textbf{\bibinfo{volume}{100}},
  \bibinfo{pages}{190404} (\bibinfo{year}{2008}).

\bibitem[{\citenamefont{Smolyarenko}(2008)}]{smo08}
\bibinfo{author}{\bibfnamefont{I.~E.} \bibnamefont{Smolyarenko}},
  \bibinfo{journal}{Phys. Rev. E} \textbf{\bibinfo{volume}{78}},
  \bibinfo{pages}{066218} (\bibinfo{year}{2008}).

\bibitem[{\citenamefont{Simons and Altshuler}(1995)}]{sim95}
\bibinfo{author}{\bibfnamefont{B.~D.} \bibnamefont{Simons}} \bibnamefont{and}
  \bibinfo{author}{\bibfnamefont{B.~L.} \bibnamefont{Altshuler}}, in
  \emph{\bibinfo{booktitle}{Proceedings of Les-Houches Summer School, session
  LXI}} (\bibinfo{address}{Les Houches}, \bibinfo{year}{1995}),
  p.~\bibinfo{pages}{81}.

\bibitem[{\citenamefont{Taniguchi et~al.}(1996)\citenamefont{Taniguchi, Simons,
  and Altshuler}}]{tan96}
\bibinfo{author}{\bibfnamefont{N.}~\bibnamefont{Taniguchi}},
  \bibinfo{author}{\bibfnamefont{B.~D.} \bibnamefont{Simons}},
  \bibnamefont{and} \bibinfo{author}{\bibfnamefont{B.~L.}
  \bibnamefont{Altshuler}}, \bibinfo{journal}{Phys. Rev. B}
  \textbf{\bibinfo{volume}{53}}, \bibinfo{pages}{R7618} (\bibinfo{year}{1996}).

\bibitem[{\citenamefont{Taniguchi et~al.}(1995)\citenamefont{Taniguchi,
  Shastry, and Altshuler}}]{tan95}
\bibinfo{author}{\bibfnamefont{N.}~\bibnamefont{Taniguchi}},
  \bibinfo{author}{\bibfnamefont{B.~S.} \bibnamefont{Shastry}},
  \bibnamefont{and} \bibinfo{author}{\bibfnamefont{B.~L.}
  \bibnamefont{Altshuler}}, \bibinfo{journal}{Phys. Rev. Lett.}
  \textbf{\bibinfo{volume}{75}}, \bibinfo{pages}{3724} (\bibinfo{year}{1995}).

\bibitem[{\citenamefont{Taniguchi et~al.}(1994)\citenamefont{Taniguchi,
  Hashimoto, Simons, and Altshuler}}]{tan94}
\bibinfo{author}{\bibfnamefont{N.}~\bibnamefont{Taniguchi}},
  \bibinfo{author}{\bibfnamefont{A.}~\bibnamefont{Hashimoto}},
  \bibinfo{author}{\bibfnamefont{B.}~\bibnamefont{Simons}}, \bibnamefont{and}
  \bibinfo{author}{\bibfnamefont{B.}~\bibnamefont{Altshuler}},
  \bibinfo{journal}{Europhys. Lett.} \textbf{\bibinfo{volume}{27}},
  \bibinfo{pages}{335} (\bibinfo{year}{1994}).

\bibitem[{\citenamefont{Dietz et~al.}(2009)\citenamefont{Dietz, Friedrich,
  Harney, Miski-Oglu, Richter, Sch{\"a}fer, Verbaarschot, and
  Weidenm{\"u}ller}}]{die09}
\bibinfo{author}{\bibfnamefont{B.}~\bibnamefont{Dietz}},
  \bibinfo{author}{\bibfnamefont{T.}~\bibnamefont{Friedrich}},
  \bibinfo{author}{\bibfnamefont{H.~L.} \bibnamefont{Harney}},
  \bibinfo{author}{\bibfnamefont{M.}~\bibnamefont{Miski-Oglu}},
  \bibinfo{author}{\bibfnamefont{A.}~\bibnamefont{Richter}},
  \bibinfo{author}{\bibfnamefont{F.}~\bibnamefont{Sch{\"a}fer}},
  \bibinfo{author}{\bibfnamefont{J.}~\bibnamefont{Verbaarschot}},
  \bibnamefont{and} \bibinfo{author}{\bibfnamefont{H.~A.}
  \bibnamefont{Weidenm{\"u}ller}}, \bibinfo{journal}{Phys. Rev. Lett.}
  \textbf{\bibinfo{volume}{103}}, \bibinfo{pages}{064101}
  (\bibinfo{year}{2009}).

\bibitem[{\citenamefont{Gorin et~al.}(2004)\citenamefont{Gorin, Prosen, and
  Seligman}}]{gor04}
\bibinfo{author}{\bibfnamefont{T.}~\bibnamefont{Gorin}},
  \bibinfo{author}{\bibfnamefont{T.}~\bibnamefont{Prosen}}, \bibnamefont{and}
  \bibinfo{author}{\bibfnamefont{T.~H.} \bibnamefont{Seligman}},
  \bibinfo{journal}{New J. Phys.} \textbf{\bibinfo{volume}{6}},
  \bibinfo{pages}{20} (\bibinfo{year}{2004}).

\bibitem[{\citenamefont{Messiah}(1999)}]{mes99}
\bibinfo{author}{\bibfnamefont{A.}~\bibnamefont{Messiah}},
  \emph{\bibinfo{title}{Quantum mechanics}} (\bibinfo{publisher}{Elsevier},
  \bibinfo{year}{1999}), \bibinfo{edition}{12th} ed.

\bibitem[{\citenamefont{Altland and Zirnbauer}(1997)}]{alt97}
\bibinfo{author}{\bibfnamefont{A.}~\bibnamefont{Altland}} \bibnamefont{and}
  \bibinfo{author}{\bibfnamefont{M.~R.} \bibnamefont{Zirnbauer}},
  \bibinfo{journal}{Phys. Rev. B} \textbf{\bibinfo{volume}{55}},
  \bibinfo{pages}{1142} (\bibinfo{year}{1997}).

\bibitem[{\citenamefont{Zirnbauer}(2011)}]{zir10}
\bibinfo{author}{\bibfnamefont{M.~R.} \bibnamefont{Zirnbauer}},
  \emph{\bibinfo{title}{Oxford Handbook on Random Matrix Theory}}
  (\bibinfo{publisher}{Oxford University Press}, \bibinfo{address}{Oxford},
  \bibinfo{year}{2011}), vol.~\bibinfo{volume}{1}, chap.~\bibinfo{chapter}{3}.

\bibitem[{\citenamefont{St{\"o}ckmann and Kohler}(2006)}]{sto06}
\bibinfo{author}{\bibfnamefont{H.~J.} \bibnamefont{St{\"o}ckmann}}
  \bibnamefont{and} \bibinfo{author}{\bibfnamefont{H.}~\bibnamefont{Kohler}},
  \bibinfo{journal}{Phys. Rev. E} \textbf{\bibinfo{volume}{73}},
  \bibinfo{pages}{066212} (\bibinfo{year}{2006}).

\bibitem[{\citenamefont{Guhr et~al.}(1998)\citenamefont{Guhr,
  M{\"u}ller-Gr{\"o}ling, and Weidenm{\"u}ller}}]{guh98}
\bibinfo{author}{\bibfnamefont{T.}~\bibnamefont{Guhr}},
  \bibinfo{author}{\bibfnamefont{A.}~\bibnamefont{M{\"u}ller-Gr{\"o}ling}},
  \bibnamefont{and} \bibinfo{author}{\bibfnamefont{H.~A.}
  \bibnamefont{Weidenm{\"u}ller}}, \bibinfo{journal}{Phys. Rep.}
  \textbf{\bibinfo{volume}{299}}, \bibinfo{pages}{189} (\bibinfo{year}{1998}).

\bibitem[{\citenamefont{St{\"o}ckmann and Sch{\"a}fer}(2004)}]{sto04}
\bibinfo{author}{\bibfnamefont{H.~J.} \bibnamefont{St{\"o}ckmann}}
  \bibnamefont{and}
  \bibinfo{author}{\bibfnamefont{R.}~\bibnamefont{Sch{\"a}fer}},
  \bibinfo{journal}{New J. Phys.} \textbf{\bibinfo{volume}{6}},
  \bibinfo{pages}{199} (\bibinfo{year}{2004}).

\bibitem[{\citenamefont{Prosen and {\v{Z}nidari\v{c}}}(2005)}]{pro05}
\bibinfo{author}{\bibfnamefont{T.}~\bibnamefont{Prosen}} \bibnamefont{and}
  \bibinfo{author}{\bibfnamefont{M.}~\bibnamefont{{\v{Z}nidari\v{c}}}},
  \bibinfo{journal}{Phys. Rev. Lett.} \textbf{\bibinfo{volume}{94}},
  \bibinfo{pages}{044101} (\bibinfo{year}{2005}).

\bibitem[{\citenamefont{St{\"o}ckmann and Sch{\"a}fer}(2005)}]{sto05}
\bibinfo{author}{\bibfnamefont{H.-J.} \bibnamefont{St{\"o}ckmann}}
  \bibnamefont{and}
  \bibinfo{author}{\bibfnamefont{R.}~\bibnamefont{Sch{\"a}fer}},
  \bibinfo{journal}{Phys. Rev. Lett.} \textbf{\bibinfo{volume}{94}},
  \bibinfo{pages}{244101} (\bibinfo{year}{2005}).

\bibitem[{\citenamefont{Simons et~al.}(1993)\citenamefont{Simons, Lee, and
  Altshuler}}]{sim93a}
\bibinfo{author}{\bibfnamefont{B.~D.} \bibnamefont{Simons}},
  \bibinfo{author}{\bibfnamefont{P.~A.} \bibnamefont{Lee}}, \bibnamefont{and}
  \bibinfo{author}{\bibfnamefont{B.~L.} \bibnamefont{Altshuler}},
  \bibinfo{journal}{Phys. Rev. B} \textbf{\bibinfo{volume}{48}},
  \bibinfo{pages}{11450} (\bibinfo{year}{1993}).

\end{thebibliography}

\end{document}